# Pressure-induced superconductivity in polycrystalline La$_3$Ni$_2$O$_{7-\delta}$

G. Wang[1,2#], N. N. Wang[1,2#*], J. Hou[1,2], L. Ma[1,3,4], L. F. Shi[1,2], Z. A. Ren[1,2], Y. D. Gu[1,2], X. L. Shen[5], H. M. Ma[5], P. T. Yang[1,2], Z. Y. Liu[1,2], H. Z. Guo[3,4], J. P. Sun[1,2], G. M. Zhang[6], J. Q. Yan[7], B. S. Wang[1,2], Y. Uwatoko[5], and J.-G. Cheng[1,2*]

[1]*Beijing National Laboratory for Condensed Matter Physics and Institute of Physics, Chinese Academy of Sciences, Beijing 100190, China*

[2]*School of Physical Sciences, University of Chinese Academy of Sciences, Beijing 100190, China*

[3]*Key Laboratory of Materials Physics, Ministry of Education, School of Physics and Microelectronics, Zhengzhou University, Zhengzhou 450052, China*

[4]*Institute of Quantum Materials and Physics, Henan Academy of Sciences, Zhengzhou 450046, China*

[5]*Institute for Solid State Physics, University of Tokyo, Kashiwa, Chiba 277-8581, Japan*

[6]*State Key Laboratory for Low dimensional Quantum Physics, Department of Physics, Tsinghua University, Beijing 100084, China*

[7]*Materials Science and Technology Division, Oak Ridge National Laboratory, Oak Ridge, Tennessee 37831, USA*

# These authors contribute equally to this work.
*Corresponding authors: nnwang@iphy.ac.cn; jgcheng@iphy.ac.cn

## Abstract

We synthesized polycrystalline La$_3$Ni$_2$O$_{7-\delta}$ ($\delta \approx 0.07$) samples by using the sol-gel method without post-annealing under high oxygen pressure, and then measured temperature-dependent resistivity under various hydrostatic pressures up to 14.5 GPa in a cubic anvil cell apparatus. We find that the density-wave-like anomaly in resistivity is progressively suppressed with increasing pressure and the resistivity drop corresponding to the onset of superconductivity emerges at pressure as low as ~7 GPa. Zero resistivity is achieved at 9 GPa below $T_c^{zero} \approx 6.6$ K, which increases quickly with pressure to 35.6 K at 14.5 GPa. The observation of zero-resistance state in the polycrystalline La$_3$Ni$_2$O$_{7-\delta}$ samples under high pressures not only corroborates the recent report of superconductivity in the pressurized La$_3$Ni$_2$O$_7$ crystals but also facilitates further studies on this emerging family of nickelate high-$T_c$ superconductors.

# Introduction

High-$T_c$ superconductors have been at the forefront of scientific exploration due to their immense potential for transformative technological applications. The groundbreaking discovery of cuprates high-$T_c$ superconductors [1,2], where superconductivity emerges through doping Mott insulators with strong electron correlations [3,4], has motivated numerous endeavors in the past decades to unveil its mechanism and to find more superconducting families with high $T_c$. Through sharing striking structural and electronic similarities with cuprates, the nickelates with $Ni^+(3d^9)$ electron configuration offer a tantalizing avenue for uncovering new high-$T_c$ superconductors. However, superconductivity was not experimentally realized in nickelates until 2019, when the infinite-layer $Nd_{1-x}Sr_xNiO_2$ thin films was found to show superconductivity with $T_c$ around 9-15 K [5]. Then, considerable dedication has been directed toward finding more nickelate superconductors with higher $T_c$[6,7]. It was later shown that the $T_c$ of $Pr_{0.82}Sr_{0.18}NiO_2$ thin films can be enhanced to over 30 K at 12.1 GPa [8]. However, the superconductivity observed in the nickelate thin films ceases to appear in the bulk samples [9].

Very recently, Sun et al. reported the observation of high-temperature superconductivity in $La_3Ni_2O_7$ crystals with $T_c$ up to 80 K under pressures above 14 GPa [10]. In contrast to the infinite-layer $Nd_{1-x}Sr_xNiO_2$, $La_3Ni_2O_7$ exhibits an exceptionally unique electronic configuration with the nominal oxidation state of $Ni^{2.5+}$, which can be considered as a mixed valence state of $Ni^{2+}(3d^8)$ and $Ni^{3+}(3d^7)$. According to the structural study under high pressure, a structural phase transition from the orthorhombic *Amam* to *Fmmm* space group occurs at about 14-15 GPa, where the interlayer Ni-O-Ni bond angle changes from 168° to 180°[10]. Subsequent high-pressure studies on $La_3Ni_2O_7$ crystals confirmed the presence of a zero-resistance state under better hydrostatic pressure conditions [11,12], in support of the discovery of a new family of nickelate high-$T_c$ superconductors. Such a remarkably high $T_c$ has immediately ignited widespread theoretical investigations on the mechanism of high-temperature superconductivity [13-18]. The significance of interlayer exchange between the $d_{z^2}$ orbitals and intra-layer hybridization of the $d_{z^2}$ and $d_{x^2-y^2}$ orbitals on the nearest neighbor sites has received substantial attention[19]. In contrast to the extensive theoretical investigations, experimental progress appears to have lagged behind, presumably due to the challenges associated with obtaining high-quality $La_3Ni_2O_7$ single crystals with controlled and homogeneous stoichiometry. Depending on the post-annealing process, the oxygen content of $La_3Ni_2O_7$ can vary from $O_{6.35}$ to $O_{7.05}$. In addition, other competitive Ruddlesden-Popper phases are commonly observed in the crystals grown using the optical image floating-zone furnace under moderate oxygen pressures. It thus becomes an important issue to perform a comprehensive study on the samples with well-controlled quality. Additionally, an open question remains concerning whether superconductivity can be achieved in $La_3Ni_2O_7$ polycrystalline samples subjected to high pressure. Therefore, we are motivated to prepare phase-pure polycrystalline $La_3Ni_2O_{7-\delta}$ samples in which oxygen content and chemical homogeneity can be easily controlled, and then to study the pressure effects on its electrical transport properties under high pressure.

In this work, we synthesized high-quality $La_3Ni_2O_{7-\delta}$ polycrystalline samples with sol-gel method and then performed a comprehensive study on the transport properties by using the piston-cylinder cell (PCC) and cubic anvil cell (CAC) under various hydrostatic pressures up to 14.5 GPa. We observed superconductivity in the pressurized $La_3Ni_2O_{6.93}$ polycrystalline samples, which exhibit zero-resistance behavior in a relatively large pressure range 9-14.5 GPa with the superconducting transition temperature $T_c^{zero}$ up to 35.6 K and $T_c^{onset}$ up to 72.2 K at 14.5 GPa. Our results show that the high-temperature superconductivity in the pressurized $La_3Ni_2O_7$ crystals can also be achieved in the $La_3Ni_2O_{6.93}$ polycrystalline samples under relatively lower pressures. The constructed $T$-$P$ phase diagram reveals the close relationship between superconductivity, density-wave-like order and the strange metal behavior.

**Experimental**

Polycrystalline $La_3Ni_2O_{7-\delta}$ samples were synthesized by using the sol-gel method and post-sintering treatment. Stoichiometric amount of $La_2O_3$ (Alfa, 99.99%) and $Ni(NO_3)_2 \cdot 6H_2O$ (Alfa, 99.99%) were dissolved in nitric acid. After adding some citric acid, the mixture was continuously stirred in a 90 °C water bath for approximately 4 hours, resulting in the formation of a vibrant green nitrate gel. This gel was then subjected to overnight heat treatment at 150-200 °C, leading to the formation of a fluffy yellow product. Afterward, the product underwent a pre-sintering step at 800°C for 6 hours to eliminate excess organic components. Subsequently, the resulting powder, with a blackish-gray appearance, was ground and pressed into pellets. These pellets were further sintered in an air environment at temperatures ranging from 1100 to 1150°C for a duration of 24 hours, ultimately yielding phase-pure polycrystalline $La_3Ni_2O_{7-\delta}$ samples.

The powder X-ray diffraction (XRD) data were collected at room temperature by PANalytical X'Pert PRO with a rotating anode (Cu $K_\alpha$, $\lambda$=1.5406 Å). The structural parameters were extracted via refining the XRD pattern with the Rietveld method using the FULLPROF program. Thermogravimetric analysis (TGA) measurement was accomplished in NETZSCH STA 449F3, using a 10% $H_2$/Ar gas flow of 50 mL/min with a 7.5 °C/min rate up to 750°C. The chemical composition and microstructure analysis were performed on a Hitachi model S-4800 field emission scanning electron microscope (SEM) with an energy-dispersive spectrometer (EDS). Temperature-dependent resistivity $\rho(T)$ at ambient pressure was measured using a Quantum Design

Physical Properties Measurement System (QD-PPMS) from 2 K to 300 K.

We employ the PCC and palm-type CAC to measure $\rho(T)$ of $La_3Ni_2O_{7-\delta}$ polycrystalline samples under various hydrostatic pressures up to 2.23 GPa and 14.5 GPa, respectively. The resistivity was measured with the standard four-probe method. Daphne 7373 and glycerol were employed as the liquid pressure transmitting medium (PTM) in PCC and CAC, respectively. The pressure values inside the PCC were estimated by measuring the $T_c$ of Sn according to the equation: $P$ (GPa) = $(T_0 - T_c)/0.482$, where $T_0 = 3.72$ K is the $T_c$ of Sn at ambient pressure. The pressure values inside the CAC were estimated from the pressure-loading force calibration curve determined by measuring the structure phase transitions of Bi, Sn and Pb at room temperature. As shown in our previous work, the three-axis compression geometry together with the adoption of liquid PTM ensures excellent hydrostatic pressure conditions up to 14.5 GPa in CAC [27]. All low-temperature measurements were performed by a $^4$He refrigerated cryostat equipped with a 9 T superconducting magnet.

**Results and discussion**

Figure 1 shows the XRD pattern of the synthesized $La_3Ni_2O_{7-\delta}$ samples. The Rietveld refinement confirms that we obtained a single-phase sample with an orthorhombic structure (space group *Amam*, No. 63). As illustrated in Fig. 1(a), the refinements converged well with reliable factors $R_p$ = 2.76 %, $R_{exp}$ = 2.58 % and $\chi^2$ = 1.92. The obtained lattice parameters shown in Fig. 1(a) are in good agreement with those reported previously [10,20-23]. To determine the oxygen stoichiometry of this compound, we performed TGA measurement in a 10% $H_2$/Ar flow. As shown in Fig. 1(b), the reduction of $La_3Ni_2O_{7-\delta}$ occurs in two steps with a final formation of a mixture of $La_2O_3$ and Ni (confirmed by Powder XRD). The oxygen stoichiometry of this phase was determined as $La_3Ni_2O_{6.93(1)}$ by calculating the weight loss between the initial and final products. Our EDX analysis confirms the chemical composition is La:Ni = 3.02(4):2 when setting Ni as 2, which is very close to the expected stoichiometry, and the EDX elemental mapping verifies the uniform distribution of these elements, as seen in Fig. 1(c, d).

Figure 2(a) shows the $\rho(T)$ of $La_3Ni_2O_{6.93}$ polycrystalline sample #1 under various pressures up to 2.23 GPa by using the PCC. At ambient pressure (AP), the $\rho(T)$ exhibits a weaker temperature dependence at high temperatures with a broad hump around 220 K, and then displays a metal-insulator-like transition behavior at $T_{dw} \approx 140$ K. The observed "weak insulating" $\rho(T)$ of $La_3Ni_2O_{6.93}$ polycrystalline sample is similar to the previously reported experimental results and the metal-insulator-like transition has been attributed to a density wave (DW) transition [20,24-26]. As shown in Fig. 2(a), the evolution of the DW transition with pressure can be tracked from the resistivity anomaly.

As the pressure gradually increases, the anomaly in $\rho(T)$ and the corresponding $T_{dw}$ determined from the minimum of $\rho(T)$ around the transition continuously moves to lower temperatures and eventually reaches about 103 K at 2.23 GPa. In addition, the DW-like characteristic undergoes broadening as pressure increases, suggesting that the long-range-ordered DW state is partially disrupted by the applied pressure.

To further track the evolution with pressure of DW transition and check whether superconductivity can be induced in the $La_3Ni_2O_{7-\delta}$ polycrystalline samples, we perform the resistivity measurements at the higher pressure range by employing CAC. Figure 2(a) and (b) displays the obtained $\rho(T)$ data of sample #2 which is prepared in the same batch with the sample #1, under various pressures up to 14.5 GPa in CAC. At AP, the $\rho(T)$ data exhibits the same behavior as observed in sample #1, confirming that our $La_3Ni_2O_{6.93}$ samples are uniform, as verified by the EDS elemental mapping. As shown in Fig. 2(a), with increasing pressure to about 3 GPa in CAC, the DW transition temperature reaches about 90 K. As the pressure continues to increase, it becomes hard to define due to the broadening of the anomalous feature in resistivity. Interestingly, a distinct behavior characterized by a resistivity drop below 33.8 K emerges at 7 GPa, and this behavior becomes more pronounced shifting to 54.2 K as the pressure is increased to 8 GPa. It is worth noting that this behavior also exhibits sensitivity to external magnetic fields. This feature motivated us to measure $\rho(T)$ in a finer pressure interval from 9 to 14.5 GPa. When the pressure approaches 9 GPa, the shallow minimum feature in $\rho(T)$ fades away and zero resistance is observed at $T_c^{zero}$= 6.6 K, signaling the occurrence of superconducting (SC) transition. This critical transformation between different electronic orders suggests that the DW order and superconductivity competes with each other. Upon further compression, the onset of the superconducting transition $T_c^{onset}$ increases slowly from 63.33 K at 9 GPa to 72.2 K at 14.5 GPa while the zero-resistance temperature $T_c^{zero}$ increases rapidly from 6.6 K at 9 GPa to 35.6 K at 14.5 GPa. Considering that the short-range DW order may partially exist, it may exhibit an inhibitory effect on superconductivity and will result in a broad SC transition. The observed different pressure dependences of $T_c^{onset}$ and $T_c^{zero}$ may associate with the competitive relationship between superconductivity and DW. Furthermore, with the enhancement of superconductivity, the $\rho(T)$ in the normal state exhibits a linear-temperature-dependence of strange metal behavior above 13.5 GPa, as shown by the solid line in Fig. 2(b). This observation suggests that the occurrence of strange metal behavior is related to the high-temperature superconductivity in the $La_3Ni_2O_{6.93}$ polycrystalline samples, which is consistent with the previous report for the $La_3Ni_2O_7$ crystal [10,11].

To further determine the observed resistance drop is truly associated with a superconducting transition, we performed detailed $\rho(T)$ measurements at 14.5 GPa under various magnetic fields. As displayed in Fig. 3(a), the superconducting transition of $La_3Ni_2O_{6.93}$ is gradually suppressed to lower temperatures and the transition width becomes broader with increasing magnetic field. Here we define $T_c^{90\%}$ and $T_c^{50\%}$ at each field according to the criteria of 90% and 50% of the corresponding normal-state resistance at $T_c^{onset}$ and plotted the temperature dependence of $\mu_0 H_{c2}(T_c)$ in Fig. 3(b). By using the empirical Ginzburg-Landau equation, the zero-temperature-limit upper critical field were determined as $\mu_0 H_{c2}(0)$ = 86.6 T and 19.1 T for $T_c^{90\%}$ and $T_c^{50\%}$, respectively.

Based on the above high-pressure characterizations, we construct the temperature-pressure (T-P) phase diagram of $La_3Ni_2O_{6.93}$ polycrystalline samples, as shown in Fig. 4. In the low-pressure region, the $La_3Ni_2O_{6.93}$ polycrystalline samples exhibit "weak insulating" behavior with a DW-like transition. As the pressure increases, the DW transition is gradually suppressed from $T_{dw} \approx 140$ K at AP to $T_{dw} \approx 90$ K at 3 GPa, above which the DW-like feature fades away and is replaced by a broad minimum centered around 90 K in resistivity. Such a shallow-valley feature vanishes completely at 9 GPa and the zero-resistance state appears concomitantly. Upon further increasing pressure, the superconducting transition temperature $T_c^{zero}$ increases rapidly from 6.6 K at 9 GPa to 35.6 K at 14.5 GP and the onset of the superconducting transition $T_c^{onset}$ reaches 72.2 K at 14.5 GPa. In addition, in concomitant with the enhancement of superconductivity in this pressure range, the strange metal behavior is observed above 13.5 GPa. As can be seen from the phase diagram, the evolution of superconductivity in the $La_3Ni_2O_{6.93}$ polycrystalline samples under high pressure shares the same trend as that observed in the $La_3Ni_2O_7$ crystals [10-12] but the critical pressure for the emergence of superconductivity in the $La_3Ni_2O_{6.93}$ polycrystalline samples is much lower. All these observations indicate that high-temperature superconductivity can also be achieved in the pressurized $La_3Ni_2O_{6.93}$ polycrystalline samples and our results reveal the close relationship between superconductivity, DW order and the strange metal behavior.

## Conclusion

In summary, phase-pure polycrystalline samples of $La_3Ni_2O_{7-\delta}$ with slight oxygen deficiency were prepared via the sol-gel method without additional oxygen annealing. Such sample exhibits a semiconducting-like electrical transport behavior with a clear upturn at $T_{dw} \approx 140$ K associated with the DW transition. Measurements of the resistivity under various hydrostatic pressures up to 14.5 GPa show that the DW related anomaly in resistivity is suppressed gradually by pressure and the superconductivity can emerge at pressures as low as ~7 GPa. The superconducting transition temperature increases progressively with further increasing pressure, reaching $T_c^{onset}$ = 72.2 K and $T_c^{zero}$ = 35.6 K at 14.5 GPa. The constructed T-P phase diagram of $La_3Ni_2O_{6.93}$ polycrystalline samples shares similar features with that of $La_3Ni_2O_7$ crystals and reveals the close relationship between superconductivity, DW order and the strange metal behavior in this system. It's noteworthy that the critical pressure required for the onset of superconductivity in $La_3Ni_2O_{6.93}$ polycrystalline samples is significantly lower than that in single crystals. Moreover, the achievement of $T_c^{zero}$ = 35.6 K at 14.5 GPa for polycrystalline samples surpasses $T_c^{zero}$ ~ 30 K at the same pressure for single crystals. Given the relative ease of synthesizing uniform polycrystalline samples with controlled stoichiometry, further investigations of such samples may provide valuable insights into the underlying factors responsible for the high-temperature superconductivity observed in $La_3Ni_2O_7$.

## Acknowledgments

J.-G. Cheng is grateful to Prof. Zhongxian Zhao for the insightful discussions and also thanks Prof. Meng Wang for the previous collaboration on the $La_3Ni_2O_7$ crystals. This work is supported by the National Natural Science Foundation of China (12025408, 11921004, 11888101), National Key Research and Development Program of China (2021YFA1400200), the Beijing Natural Science Foundation (Z190008), the Strategic Priority Research Program of CAS (XDB33000000), the Specific Research Assistant Funding Program of CAS (E3VP011X61) and the Users with Excellence Program of Hefei Science Center CAS (2021HSC-UE008). J.Q.Y was supported by the US Department of Energy, Office of Science, Basic Energy Sciences, Materials Sciences and Engineering Division. The high-pressure experiments were performed at the Cubic Anvil Cell station of Synergic Extreme Condition User Facility (SECUF).

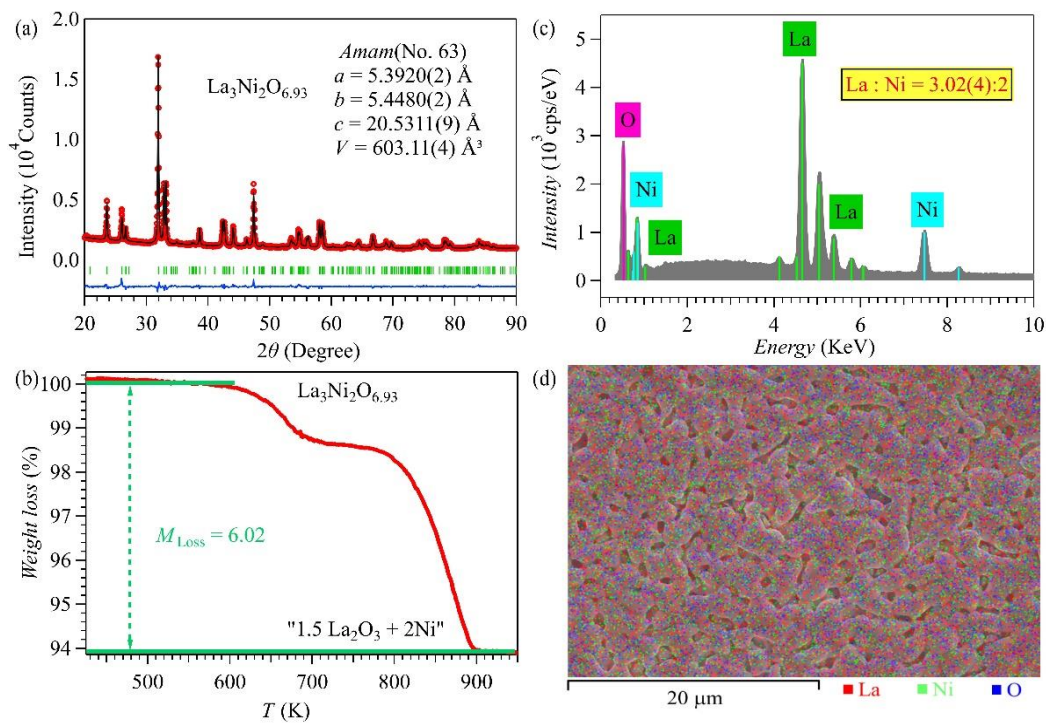

Fig 1. (a) Rietveld refinements on the XRD pattern of $La_3Ni_2O_{7-\delta}$ polycrystalline sample. The obtained lattice parameters are shown in the figure. The bottom marks and line correspond to the calculated Bragg diffraction positions and the difference between observed and calculated data, respectively. (b) Thermogravimetric curves for $La_3Ni_2O_{7-\delta}$ in 10% $H_2$/Ar. (c) and (d) The SEM-EDS elemental mapping of $La_3Ni_2O_{7-\delta}$ polycrystalline sample.

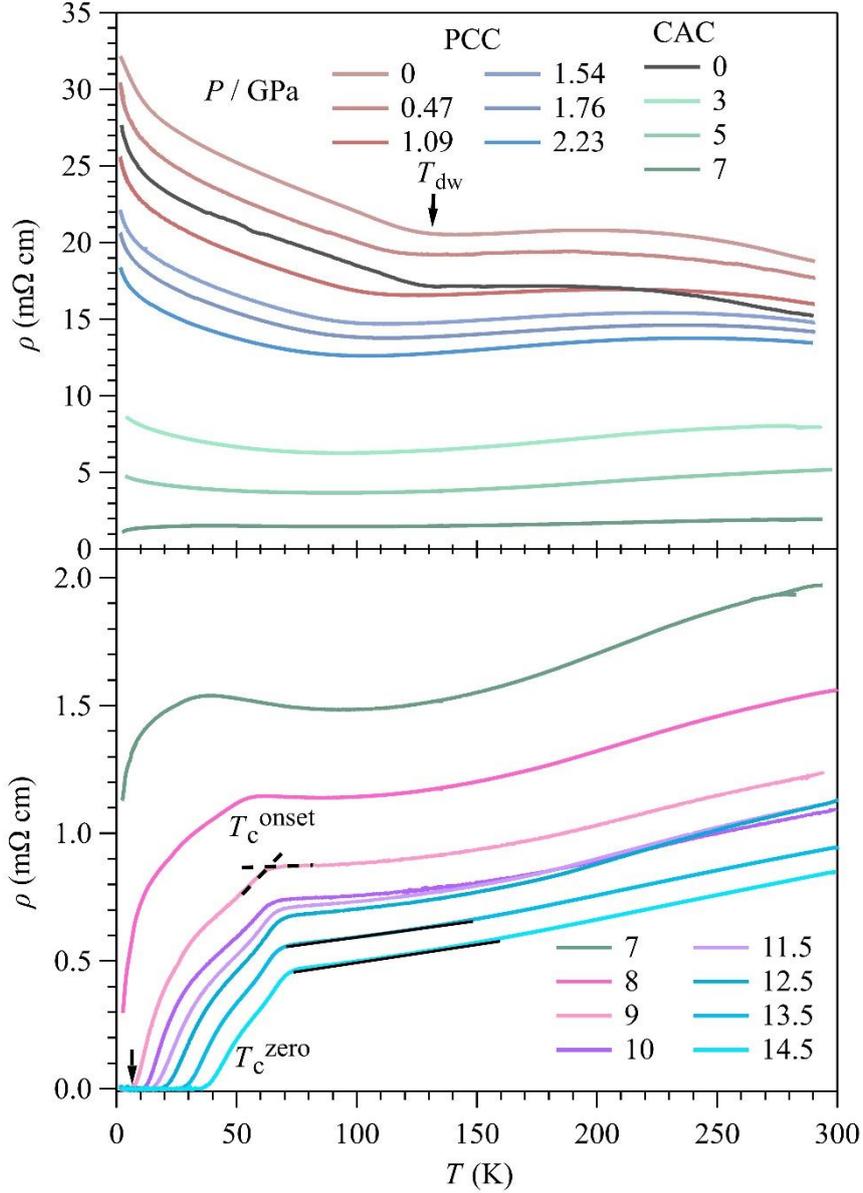

Fig 2. (a) Temperature dependence of resistivity $\rho(T)$ of La$_3$Ni$_2$O$_{6.93}$ polycrystalline samples under various hydrostatic pressures up to 14.5 GPa measured in PCC and CAC. Here, the $T_c^{onset}$ is determined as the interception between two straight lines below and above the superconducting transitions. (b) The enlarged view of low temperature $\rho(T)$ data at pressures from 7 to 14.5 GPa, highlighting the gradual development of the resistance drop upon compression.

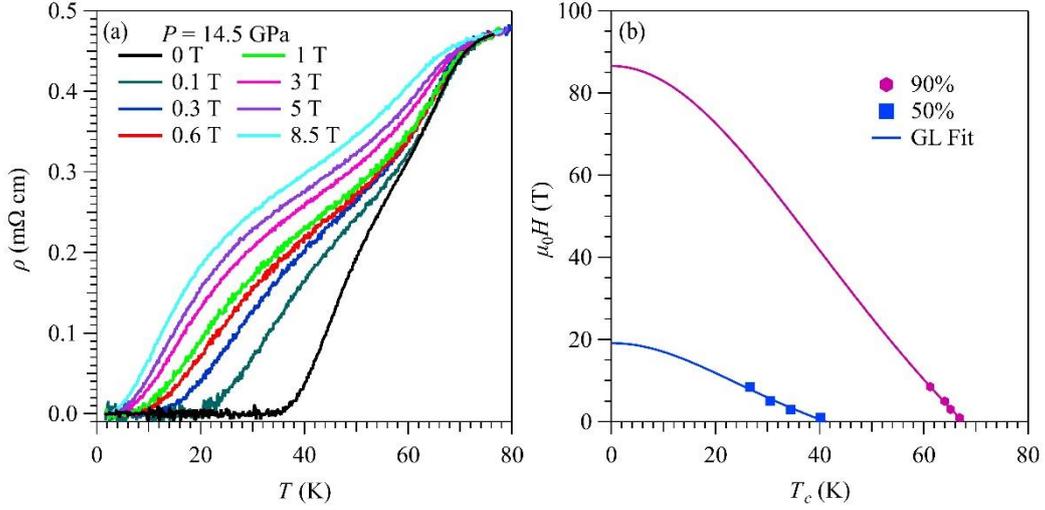

Fig 3. (a) The low-temperature resistivity $\rho(T)$ at 14.5 GPa under various magnetic fields up to 8.5 T. (b) Temperature dependence of the upper critical field $\mu_0H_{c2}(T)$ for La$_3$Ni$_2$O$_{6.93}$ polycrystalline sample at 14.5 GPa. The solid line is the fitting curve by using the formula $H_{c2}= H_{c2}(0)(1-t^2)/(1+t^2)$, where $t = T/T_c$.

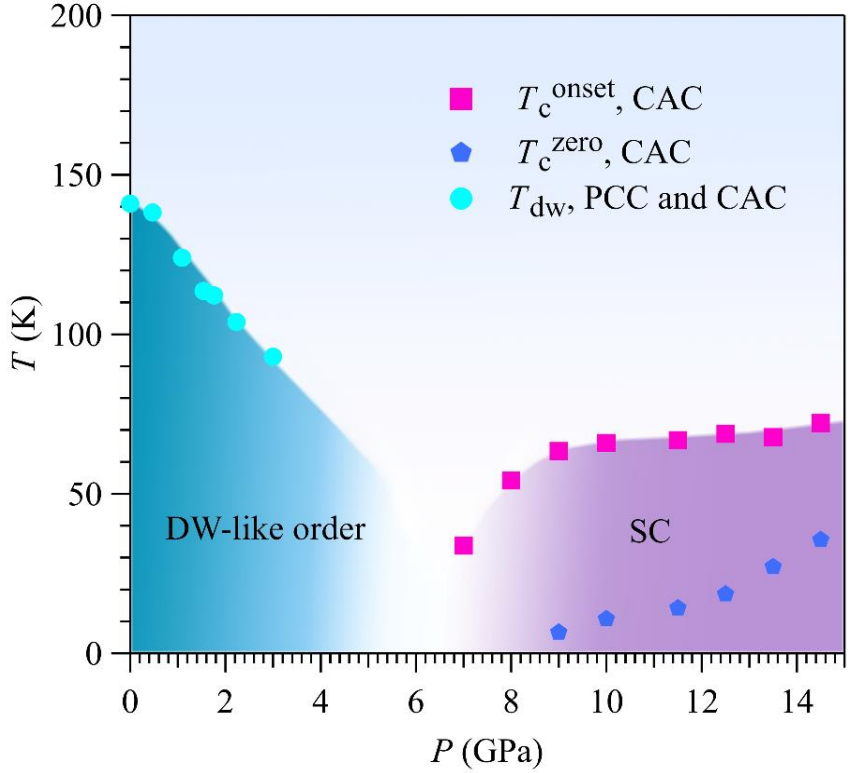

Fig 4. The $T$-$P$ phase diagram of the La$_3$Ni$_2$O$_{6.93}$ polycrystalline sample. The solid circles represent the DW-like transition $T_{dw}$ measured at various pressures using PCC and CAC, respectively. The solid squares and pentagons represent the onset and zero-resistance superconducting transition temperatures determined from the present measurements in CAC.